\documentclass[aps,prl,twocolumn,showpacs,groupedaddress]{revtex4}
\usepackage{graphicx}
\begin{document}
\title{Cooperative electron-phonon interaction in  molecular chains}
\author{ Sudhakar Yarlagadda}
\affiliation{Cavendish Lab, Univ. of Cambridge, UK and 
Saha Institute of Nuclear Physics, Calcutta, India}
\date{\today}
\begin{abstract}
Using a controlled analytic treatment, we derive a
model that generically describes cooperative 
strong electron-phonon interaction (EPI) in one-band 
 and two-band Jahn-Teller (JT) systems. The model 
involves a {\em next-nearest-neighbor} hopping
and a nearest-neighbor repulsion.
Rings with odd number of sites (o-rings) belong to a different universality
class compared to rings with even number (e-rings). The 
e-rings, upon tuning repulsion,
 undergo a dramatic discontinuous transition
to a {\em conducting commensurate}
charge density wave (CDW) state with a period independent
of filling.
\end{abstract}
\pacs{
 71.38.-k, 71.45.Lr, 71.38.Ht, 75.47.Lx  } 

\maketitle
The last few decades have witnessed numerous
 studies to fathom the tapestry of
exotic  orbital, charge, and spin orderings in JT
 transition metal oxides (TMO)
such as the manganites, cuprates, nickelates, etc. \cite{hotta}.
To model the emergent ordering in these complex TMOs, one needs,
as building blocks,
effective Hamiltonians for various interactions.
Except for the cooperative JT interaction,
effective Hamiltonians, that reasonably mimic the physics,
 have been derived for all other interactions. For instance,
double exchange model approximates infinite Hund's coupling, Gutzwiller
approximation or dynamical mean-field theory model
 Hubbard on-site coulombic interaction, superexchange
describes localized spin interaction at strong on-site repulsion, etc. 
Controlled mathematical modeling of cooperative JT quantum systems 
that goes beyond modeling localized carriers  \cite{gehring}
has remained elusive (at least to our knowledge).
In fact, a controlled analytic treatment of the many-polaron effects
in single-band  Holstein model \cite{holstein}
(which is a simpler non-cooperative EPI system) has 
been reported only recently \cite{sdadys}.
However, definite progress has been made in numerically treating
JT systems \cite{hotta}.

Owing to its cooperative nature, the JT interaction leads to non-local
distortion effects which can change the very nature of long range order.
The situation is further complicated by the ubiquitous strong EPI
for which there is compelling evidence in manganites (from
 extended X-ray absorption fine structure \cite{bianc} and
 pulsed neutron diffraction \cite{louca} measurements),
in cuprates (through angle-resolved photoemission spectroscopy
 \cite{damascelli}), and in nickelates (based
on neutron scattering \cite{tranquada}).
While a weak interaction is amenable to a Migdal-type  of perturbative
treatment, the strong interaction (even for a single-band system)
necessitates a non-perturbative approach \cite{sdadys}.

\begin{figure}[t]
\includegraphics[width=3.0in]{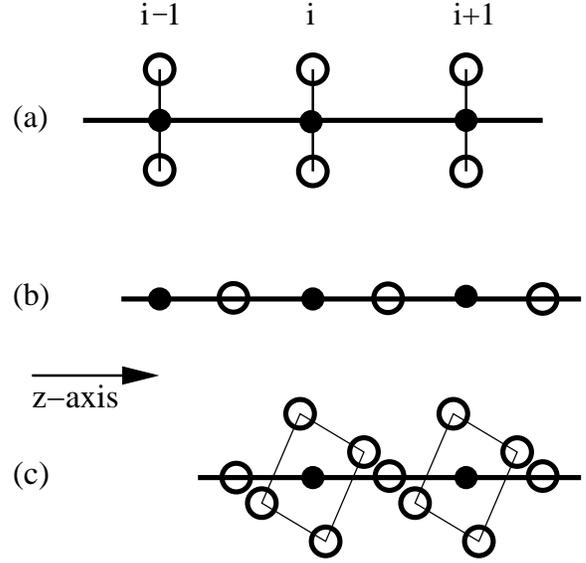}
\noindent\caption[]{Molecular chains with  $d_{z^2}$ orbital hopping sites
(filled circles) and oxygen sites (empty circles) in
(a) Holstein model,
(b) one-band cooperative EPI system, 
 and (c) two-band cooperative JT system with $Q_1$, $Q_2$, and $Q_3$ octahedral
modes.}
\label{chain}
\end{figure}

In this letter, based on our
work \cite{sdadys}
on the Holstein model [Fig. \ref{chain}(a)],
we derive the effective Hamiltonian for cooperative strong EPI
in one-band system [Fig. \ref{chain}(b)] and two-band (from $d_{z^2}$ and
$d_{x^2-y^2}$ orbitals) JT chain [Fig. \ref{chain}(c)].
 Upon inclusion of cooperative effects in the strong EPI,
the system changes its dominant transport mechanism from
one of nearest-neighbor hopping to that of next-nearest-neighbor hopping.
Below half-filling and at strong EPI in e-rings,
while the 
systems without cooperative effects
remain disordered Luttinger liquids, our cooperative EPI one-band
and two-band JT systems 
produce 
CDW states.
Furthermore, 
{\it like the 
 manganites,
 our JT system too manifests absence of electron-hole symmetry only upon
including cooperative effects}.

To bring out the essential physics, we begin with a one-dimensional
(1D) toy-system of electrons
 hopping in a one-band system of $d_{z^2}$ orbitals which are coupled to 
the oxygens in between as shown in Fig. \ref{chain}(b) \cite{sawatzky}.
The Hamiltonian is expressed as
 $H = H_t + H_{ep} + H_l$
where the hopping term $H_t$ is given by
\begin{eqnarray}
H_t = 
- t \sum_{j} ( c^{\dagger}_{j} c_{j+1} + {\rm H.c.}) , 
\label{Ht}
\end{eqnarray}
with $c_j$ being the destruction
 operator of an electron in a $d^j_{z^2}$ orbital (at site $j$)
and the EPI term $H_{ep}$ by
\begin{eqnarray}
H_{ep} = 
- g \omega_0 \sqrt{2 M \omega_0} 
 \sum_{j}  n_j
 q_j , 
\end{eqnarray}
with 
$n_i = c^{\dagger}_{i} c_{i} $,
$q_i = u_i - u_{i-1}$ representing the expansion of the oxygens around the
$d^i_{z^2}$ orbital, and
 the right-hand-side (RHS) oxygen displacement 
$u_i =(a^{\dagger}_i + a_i)/{\sqrt{2 M \omega_0}}$. Furthermore,
 the lattice term 
$H_l$ representing simple harmonic oscillators is 
of the form
\begin{eqnarray}
H_{l} = 
\frac{K}{2} \sum_{j} u_j^2   
+ \frac{1}{2M} \sum_{j} p_j^2 =  
 \omega_0 \sum_{j} a^{\dagger}_{j} a_{j} .
\end{eqnarray}
 The main difference between the Holstein model and the above
cooperative Hamiltonian is that in the Holstein model  electrons at
different sites are
coupled to  different on-site molecular distortions
whereas in the present toy-system
 the electrons on adjacent sites are coupled
to the displacement of the same in-between oxygen. 
Thus in our toy-system, to produce an effective polaronic Hamiltonian,
 we need to devise a modification of the usual Lang-Firsov
transformation \cite{LF} so as to take into account the 
cooperative nature of the
 distortions. To meet this end we used the following canonical
transformation
 $\tilde{H} = \exp(S) H \exp(-S)$
where
$S$ now contains the {\em difference in densities on adjacent sites}
\begin{eqnarray}
S = g \sum_j ( a_{j} - a^{\dagger}_{j})
(n_j - n_{j+1}) .
\end{eqnarray}
Then, one obtains
 $\tilde{H}  = H_0 + H_1$ where
\begin{eqnarray}
H_0 =&& 
 \omega_0 \sum_{j} a^{\dagger}_{j} a_{j} -
2 g^2 \omega_0
\sum_{j} n_j 
+ 2 g^2 \omega_0 \sum_{j} n_j n_{j+1} 
\nonumber \\
&& - t e^{-3 g^2} \sum_{j} ( c^{\dagger}_{j} c_{j+1} + {\rm H.c.}) , 
\end{eqnarray}
and 
\begin{eqnarray*}
H_1 = \sum_{j}H_{1j} =
- t e^{-3 g^2}
 \sum_{j} [ c^{\dagger}_{j} c_{j+1}
\{
{\cal{T}}_{+}^{j \dagger} {\cal{T}}^{j}_{-}
-1 \} + {\rm H.c.}] , 
\end{eqnarray*}
with
 ${\cal{T}}^{j}_{\pm} = \exp[\pm g( 2 a_{j} - a_{j-1} - a_{j+1})]$.
{\it On account of the cooperative nature of the EPI
we obtain an additional term 
$ 2 g^2 \omega_0 \sum_{j} n_j n_{j+1}$} 
involving nearest-neighbor repulsion in $H_0$
 and the perturbation $H_1$ now involves
phonons at three sites as opposed to phonons at only two sites as in
 the non-cooperative case.
We consider the case $ t\exp[-3 g^2] << \omega_0$ and 
perform second-order perturbation theory similar
to that in Ref. \onlinecite{sdadys}.
For large $g^2$, after tedious algebra,
we obtain the following effective polaronic Hamiltonian
 \cite{ys2}:
\begin{eqnarray}
H^{C}_{eff} = && 
-\left [ 2 g^2 \omega_0 
 +\frac{t^2}{3 g^2 \omega_0} \right ]
\sum_j n_{j+1} (1-n_{j})
\nonumber \\
&&
 - t e^{-3 g^2} \sum_{j} ( c^{\dagger}_{j} c_{j+1} + {\rm H.c.}) 
\nonumber \\
&&
 - \frac{t^2 e^{-2 g^2} }{4 g^2 \omega_0}
 [c^{\dagger}_{j-1}(1-2n_j) c_{j+1} + {\rm H.c.} ] .
\label{Hpol}
 \end{eqnarray}                               
Notice that the coefficient of the nearest-neighbor hopping is significantly
smaller than the next-nearest-neighbor hopping for
large $g^2$ and not-too-small $t/\omega_0$!
This is a {\em key feature resulting from cooperative effects.}

The above effective Hamiltonian
may be contrasted with the following  Hamiltonian $H_{eff}$
for the case where there is no cooperative EPI
 [i.e., $H_{ep} = -\sqrt{2}g \omega_0 \sum_i
n_i (a^{\dagger}_i + a_i)$]
 [see Ref. \onlinecite{sdadys} and Fig. \ref{chain}(a)]:
\begin{eqnarray}
H_{eff} = && 
-2 g^2 \omega_0 \sum_j n_j
 -\frac{t^2}{2 g^2 \omega_0} \sum_j n_{j+1} (1-n_{j})
\nonumber \\
&&
 - t e^{-2 g^2} \sum_{j} ( c^{\dagger}_{j} c_{j+1} + {\rm H.c.}) 
\nonumber \\
&&
 - \frac{t^2 e^{-2 g^2} }{2 g^2 \omega_0}
\sum_j [c^{\dagger}_{j-1}(1-2n_j) c_{j+1} + {\rm H.c.} ] .
\label{Hpolold}
 \end{eqnarray}                               
Although our results in Eq. (\ref{Hpolold}) are valid for
$t e^{- 2g^2} << \omega_0$ and $g^2 >> 1$,
we provide an explanation of the results in the more restrictive but
physically  understandable adiabatic ($t >> \omega_0$) and small polaronic 
($g^2 \omega_0 >> t$) regime as follows.  In 
Eq. (\ref{Hpolold}), the coefficient of the 
$ \sum_j n_{j+1} (1-n_{j})$
 term can be understood as resulting from an adiabatic process
 when an electron at site $j+1$ hops to a neighboring site $j$ 
and back, but the
lattice has no time to distort (relax) {\em locally}
at site $j$ ($j+1$)
and thus yields the second-order perturbation energy $-t^2$/(energy change).
On the other hand the coefficient of
$\sum_j [c^{\dagger}_{j-1}(1-2n_j) c_{j+1} + {\rm H.c.} ] $
results when the intermediate site $j$ does not distort/relax in an 
adiabatic hopping and thus yields 
$ t\exp[-2 g^2] \times \frac{t}{2 g^2 \omega_0}$ where
$ t\exp[-2 g^2] $ is due to distortions at both initial and final sites $j-1$
and  $j+1$.
In the above non-cooperative case, the nearest-neighbor hopping dominates over
the next-nearest-neighbor hopping 
in the small polaron limit.

Using the above logic we see that the
higher order terms in perturbation theory,
for both cooperative and non-cooperative cases,
 are dominated by the process where
an electron hops back and forth between the same two sites. The
dominant term to $k$th order is approximately given for even $k$ by 
\begin{eqnarray}
  \omega_0 \left [ \frac{t}{ g \omega_0} \right  ]^k 
\sum_j n_{j+1} (1-n_{j}) ,
 \end{eqnarray}                               
while for odd $k$  by
\begin{eqnarray}
t e^{-\gamma g^2}
\left [ \frac{t}{ g \omega_0} \right ]^{k-1} 
 \sum_{j} ( c^{\dagger}_{j} c_{j+1} + {\rm H.c.}) , 
 \end{eqnarray}                               
where $\gamma$ is 2 for the non-cooperative case and 3 for the cooperative one.
Since each term in the perturbation theory should be smaller than
$\omega_0$, we see that the small parameter in our perturbation
 is $t/(g \omega_0)$ \cite{sdys}.

Here, a few observations are in order.
Firstly, the cooperative effects, unlike in the Holstein
model's case, raise  
the potential of the site next to an occupied site
 and thus makes it unfavorable for hopping.
Thus, in Eq. (\ref{Hpol}) as compared to Eq. (\ref{Hpolold}),
 the exponent is larger for the nearest-neighbor hopping
and also the denominators of the coefficients 
are similarly larger for the hopping-generated
nearest-neighbor interaction
 and for the next-nearest-neighbor hopping. Next,
 Lau {\it et al.} \cite{sawatzky}
obtain the same energy expression for {\it a single polaron}
as that given by Eq. (\ref{Hpol}) (when $n_j=0$).
Lastly, in Ref. \onlinecite{tvr}, the authors explain the ferromagnetic
insulating behavior in low-doped manganites by using the
non-cooperative 
hopping-generated nearest-neighbor interaction [i.e., second term on RHS
of Eq. (\ref{Hpolold})]
after modifying the hopping term for double-exchange effects.
From Eq. (\ref{Hpol}),
 we see that cooperative phenomenon must be taken into
account as it reduces the ferromagnetism generating interaction
strength by a factor of $1.5$.

We will now analyze the effective
polaron Hamiltonian given by Eq. (\ref{Hpol}).
We first note that, for large values of $g^2$,
the essential physics is captured by 
\begin{eqnarray}
H^{C}_{eff} = &&  
 - T \sum_j (c^{\dagger}_{j-1}(1-2n_{j}) c_{j+1} + {\rm H.c.} ) 
\nonumber \\
&&
+ V \sum_j n_{j} n_{j+1} ,
\label{Hpol_model}
 \end{eqnarray}                               
with $T/V << 1$. However, owing to its novelty, we will study the
model given by Eq. (\ref{Hpol_model}) for arbitrary value of $T/V$.
Next we note that for $T/V << 1$ in 
e-rings,
 the system  always has alternate sites (one sub-lattice) occupied
for less than half-filling and above half-filling the other sub-lattice gets
filled. This is due to the fact that at large repulsion,
as we fill up the lattice with electrons  one after the other,
 each new electron added to the system has more
number of sites to hop to on the same sub-lattice occupied
by the previous electrons. As for the other extreme
situation $V=0$, for even number of electrons on e-rings,
the model has both sub-lattices equally occupied.

\begin{figure}
\includegraphics[width=3.0in]{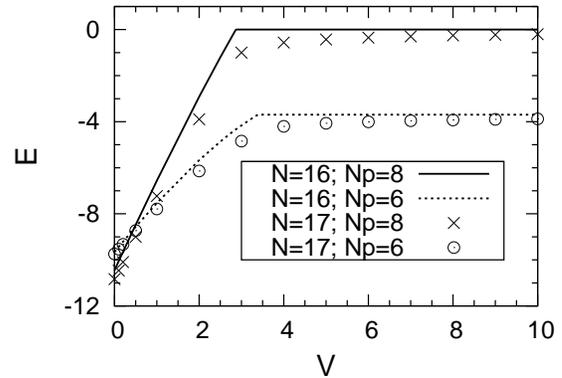}
\noindent\caption[]{Plots of Energy (E) versus interaction strength (V)
for rings with $N$ sites, $N_p$ electrons, and hopping $T=1$.}
\label{EV}
\end{figure}

\begin{figure}
\includegraphics[width=3.0in]{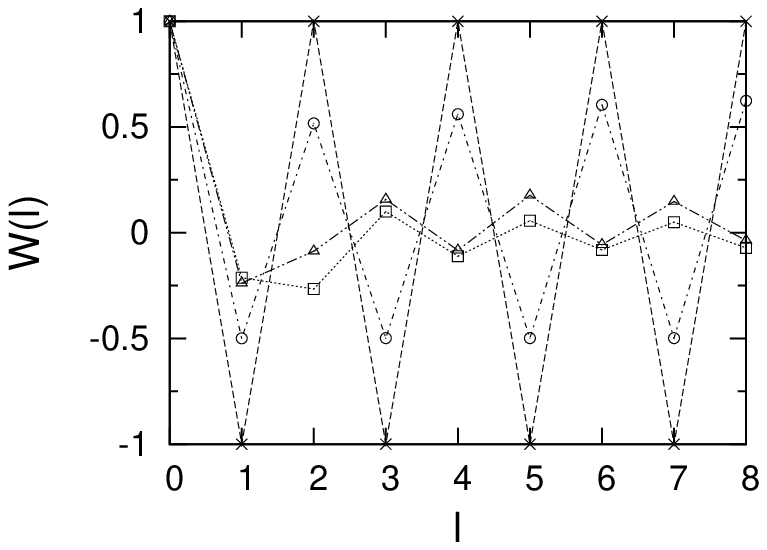}
\includegraphics[width=3.0in]{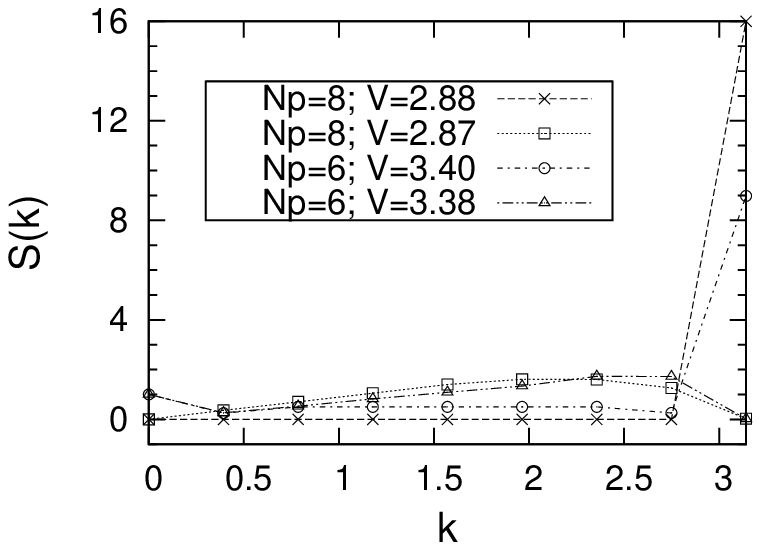}
\noindent\caption[]{Plot of the density-density correlation function
$W(l)$ and structure factor $S(k)$ for a 16-site ring with  $N_p$
electrons, interaction strength $V$, and hopping $T=1$.}
\label{wlsq}
\end{figure}

Using modified Lanczos \cite{gagliano}, we calculated
the ground state energy using periodic (antiperiodic)
boundary conditions for odd (even) number of fermions.
First {\it for e-rings with $N$ sites,
 the ground state energy has a 
slope discontinuity, with the energy increasing
up to a critical value, after which it is constant 
(see Fig. \ref{EV})}.
We will now 
show clearly that as the interaction strength increases, {\it at a critical
value of $V/T$,
Ising $Z_2$ symmetry
 (i.e., both sub-lattices being equally populated)
 is broken and only a single sub-lattice
is occupied for e-rings}. To this end, we calculate
the density-density correlation 
function $W(l) = 4/N \sum_j \langle (n_j -0.5)(n_{j+l}-0.5) \rangle$
and
 the static structure factor $S(k) = \sum_l \exp (ikl) W(l)$
near critical value.
Now, $S(\pi) = [\sum_{l {\rm even}} - \sum_{l {\rm odd}}]W(l)$
with
\begin{eqnarray}
\sum_{l {\rm even}}W(l) = 
\frac{4}{N} \{ [N_e - 0.25 N]^2 + [N_o - 0.25 N]^2 \} ,
\end{eqnarray}
and
\begin{eqnarray}
\sum_{l {\rm odd}} W(l) = \frac{8}{N} [N_e - 0.25 N] [N_o - 0.25 N] ,
\end{eqnarray}
where $N_{e(o)}$ is the number of particles on even (odd) sites.
Then, when $Z_2$ symmetry is respected,
 for even number of particles $N_p = 2N_e = 2N_o$,
we have $S(\pi)=0$ and for odd 
value of $N_p$ we have $S(\pi)= 4/N$.
When only one sub-lattice is occupied, on
taking $N_e = N_p$ and $N_o = 0$ (without loss of generality),
we get $S(\pi)=4 N_p^2/N$ and 
$W(l {\rm odd})= 1 -4 N_p/N$.
We find that 
 at a critical interaction strength,
as shown in  Fig. \ref{wlsq}, the following dramatic changes occur:
(i) the structure factor $S(\pi)$ 
jumps 
 from near $0$ to almost
its maximum value $4 N_p^2/N$; 
(ii) $W(l {\rm odd})$ also jumps close
 to its large $V$ value of $ 1 -4 N_p/N$;
and (iii)  $W(l {\rm even})$ (for $l \neq 0$) too jumps and its
final value at half-filling is 1.
For a fixed $T$ and $N$, the critical value $V^N_C$ of $V$
 increases monotonically
as $N_p$ decreases. For $N=16$, $N_p = 2$, and $T=1$,
 we get $V^{16}_{C} \approx
4$. From finite size scaling for half-filling, using 
$V^N_{C} - V^{\infty}_C \propto 1/N^2$ and system size $N \le 20$,
we obtain $V^{\infty}_C \approx 2.83$.

We see from the above analysis that, at a critical repulsion,
 the system undergoes a discontinuous
transition to a {\em conducting commensurate CDW} state away from half-filling
while at half-filling one obtains a Mott transition.
Usually commensurate CDW's are insulating (see Ref. \onlinecite{gruner})
 whereas our model surprisingly predicts a conducting commensurate CDW.
Furthermore, quite unlike the Peierls transition,
 the {\it period of the CDW is  independent of density}!
Such density independent charge ordering has indeed
been observed in manganites
(see Fig. 2 in Ref. \onlinecite{pbl}).
Our toy-system is different from the non-cooperative  EPI
Holstein model
in that the latter,
 at strong coupling and finite $t/\omega_0$,
can be mapped onto an anisotropic Heisenberg model
and is thus always a Luttinger liquid away from half-filling (see 
Ref. \onlinecite{sdadys}).

We will now discuss o-rings
for number of electrons less than half the number of sites.
The case when the number of electrons is greater can be
deduced invoking electron-hole symmetry.
Since the system has all sites connected 
through next-nearest-neighbor hopping (much like a Moebius strip),
there do not exist 2 sub-lattices. Consequently,
as the interaction increases, the system's
energy increases monotonically and 
smoothly (without any slope discontinuities) and there
is no phase transition [see Fig. \ref{EV}].
We find that the system's
energy increases linearly at small strengths 
as can be expected from a mean-filed analysis.
At large values, the system's energy approaches asymptotically the
energy of the e-ring with one less site but with the same number
of electrons. This is because at large $V$, hard core repulsion
on adjacent sites blocks electrons similar to Pauli's exclusion principle
on the same site.

Lastly, for the 1D JT system shown in Fig. \ref{chain}(c), 
we first note that there is a strong 
on-site inter-orbital repulsion
 given by 
$ U \sum_j n_j n^{\prime}_j $ where $U \rightarrow \infty$ and 
 $n^{\prime}_j =  d^{\dagger}_j d_j $ with
 $d$ being the destruction operator of a localized
electron in a $d_{x^2 - y^2}$ orbital.
Next, like the one-band case [see Eq. (\ref{Hpol})],
 the cooperative JT effect
too produces strong
nearest-neighbor replusion $\propto g^2 \omega_0 \sum_j n_j n_{j+1}$.
Again as before, 
there is a virtual-hopping generated nearest-neighbor repulsion
term 
$ - G \sum_{j, \delta = \pm 1} 
 n_{j+\delta}(1-n_j) (1 - n^{\prime}_{j}) $
with the additional 
 $ (1 - n^{\prime}_{j}) $ 
factor
prohibiting
 hopping 
of a $d_{z^2}$ electron to a neighborging
$j$ site 
if it is occupied by a 
$d_{x^2-y^2}$ electron.
Here too $G \propto t^2/(g^2 \omega_0)$.
Then,
the Hamiltonian at large EPI
(whose exact form is obtained after
 quite tedious algebra \cite{ys2})
is essentially given by
\begin{eqnarray}
H^{CJT}_{eff} = &&  
\!\!\!\!\!\!
\sum_j[V n_{j} n_{j+1} +U n_j n^{\prime}_j]  
 - G \sum_{j, \delta = \pm 1} n_{j+\delta} (1 - n^{\prime}_{j}) 
\nonumber \\
&&
\!\!\!\!\!\!
 - T \sum_j [(1-n^{\prime}_j)c^{\dagger}_{j-1}
(1-2n_{j}) c_{j+1} + {\rm H.c.} ] ,
\label{Hpol_JTmodel}
 \end{eqnarray}                               
with
$ G >> T$. 
As in Eq. (\ref{Hpol_model}),
here too $V$ is
 an arbitrary variable. 
In Eq. (\ref{Hpol_JTmodel}),
 in the next-nearest-neighbor hopping term,
the $(1-n^{\prime}_j)$ factor
 does not
allow hopping via the $j$ site if it is occupied by a $d_{x^2-y^2}$ electron.

 For fillings up to half (i.e., 0.5
electron/site), the electrons populate only the
 $d_{z^2}$ orbitals (i.e., $n^{\prime}_j = 0$) 
and the behavior of the above model
 of Eq. (\ref{Hpol_JTmodel}) is
identical to that of the effective model given by Eq. (\ref{Hpol_model})
and is depicted by Fig. \ref{EV} when $E$ is replaced by $E + 2 N_p G$.
Our results agree with the experimental
observations that 
$C$-type ferromagnetic-chain manganites,
like $Nd_{1-x} Sr_{x}MnO_3$ at $0.63 \le x \le 0.8$ \cite{tokura2},
are $d_{z^2}$ polarized.  
For fillings between half and unity and when
 $V >> G$ ({\it as in a real cooperative EPI system}), 
for e-rings [o-rings] $N/2$ [$(N-1)/2$] electrons populate
$d_{z^2}$ orbitals with the rest occupying 
$d_{x^2-y^2}$ orbitals;
thus
system lacks electron-hole symmetry.

On ignoring cooperative effects in our JT sytem of Fig. \ref{chain}(c),
in Eq. (\ref{Hpol_JTmodel})
we set $V = 2G$ 
 and add a
nearest-neighbor hopping 
$-T_0 \sum_{j, \delta = \pm 1}
c^{\dagger}_{j}c_{j+\delta}$ with $G >> T_0 >> T$ [as in Eq. (\ref{Hpolold})].
Then,
 importantly, only $d_{z^2}$ orbitals are occupied 
at any filling (up to unity)
and the system exhibits electron-hole symmetry.
Furthermore,
since $n^{\prime}_j = 0$,
the system is always a Luttinger liquid at 
non-half-filling \cite{sdadys}.

In summary, our study shows that cooperative EPI
 produces strikingly different physics
compared to the non-cooperative situation.
Our derived 
model captures
 the essential feature of cooperative
strong EPI
 in all dimensions (and in even non-cubic geometries), i.e.,
the dominant transport is due to next-nearest-neighbor hopping.
Our controlled analysis of cooperative JT
 interaction
is currently being extended to
 higher dimensions.

The author would like to thank P. B. Littlewood, S. Kos, and
 D. E. Khmelnitskii
for valuable discussions and KITP for hospitality. This work was funded
by UKIERI.


\end{document}